\documentclass{aa}  
\newcommand{\feii}{\ion{Fe}{ii}}

\newcommand{\sii}{\ion{S}{ii}}

\newcommand{\oi}{\ion{O}{i}}

\newcommand{\n}{\ion{N}{i}}

\newcommand{\nii}{\ion{N}{ii}}

\newcommand{\caii}{\ion{Ca}{ii}}

\newcommand{\Ki}{\ion{K}{i}}
\newcommand{\Lii}{\ion{Li}{i}}

\usepackage{graphicx}
\usepackage{txfonts}
\begin{document}

   \title{A new insight into the variability of V1184 Tau }






   \author{T. Giannini\inst{1}, D. Lorenzetti\inst{1}, A. Harutyunyan\inst{2}, G. Li Causi\inst{1,3}, S. Antoniucci\inst{1},  
   A. A. Arkharov\inst{4}, V. M. Larionov\inst{4,5}, F. Strafella\inst{6},
         R.Carini\inst{1}, A. Di Paola\inst{1}, R. Speziali\inst{1} 
          }

   \institute{INAF-Osservatorio Astronomico di Roma, via Frascati 33, I-00040 Monte Porzio Catone, Italy 
        \and           
           Fundaci\'{o}n Galileo Galilei - INAF, Telescopio Nazionale Galileo, 38700 Santa Cruz de la Palma, Tenerife, Spain                         
         \and
             INAF-Istituto di Astrofisica e Planetologia Spaziali, via del Fosso del Cavaliere 100, I-00133 Roma, Italy 
          \and          
          Central Astronomical Observatory of Pulkovo, Pulkovskoe shosse 65, 196140, St. Petersburg, Russia
          \and
          Astronomical Institute of St. Petersburg University, Russia            
          \and
          Dipartimento di Matematica e Fisica, Universit\'{a} del Salento, I-73100 Lecce, Italy            
          }

   \date{Received ; accepted }

 
  \abstract
{V1184 Tau is a young variable that, for a long time has been monitored at optical wavelengths. Its variability has been ascribed to a sudden and repetitive increase of the circumstellar extinction  (UXor-type variable), but the physical origin of this kind of variation, although hypothesized, has not been fully supported on an observational basis.}  
 {With the aim of getting a new insight into the variability of V1184 Tau, we present here new photometric and spectroscopic observations that were taken in the period 2008$-$2015. During that time, the source 
 reached the same high brightness level that it had before the remarkable fading of about 5 mag, which happened  in 2004. The optical spectrum is the first to be obtained when the continuum was at its maximum level.}
{All the observational data are interpreted in the framework of extinction-driven variability. In particular, we analyze light curves, optical and near-infrared colors, spectral energy distribution, and optical spectrum.}
{The emerging picture indicates that the  source fading is due to an extinction increase of $\Delta$A$_V$ $\sim$ 5 mag, which is associated with the appearance of a strong infrared excess, attributable to a 
 thermal component a T$\sim$ 1000 K. From the flux of H$\alpha,$ we derive a  mass accretion rate in the range 10$^{-11} -$ 5\, 10$^{-10}$ M$_{\odot}$ yr$^{-1}$ s, which is marginally consistent with what is expected for a classical T Tauri star of similar mass. The source spectral energy distribution was fitted for both the high- and low- level of brightness. Remarkably, a scenario that is consistent with the known stellar properties (such as spectral type, mass, and radius) is obtained only if the distance to the source is of few hundreds of parsecs, in contrast with the commonly assumed value of $\sim$ 1.5 kpc. }
 {Our analysis partially supports that presented by Grinin (2009), according to which the circumstellar disk undergoes a periodic puffing,
the observational effects of which both  shield the central star and provide evidence of disk wind activity. However, since the  mass accretion rate remains almost constant with time, the source is likely not subject to accretion bursts.} 
\keywords{Stars: formation -- stars: variables: T Tauri -- infrared: stars  --  stars: individual (V1184 Tau)}
\authorrunning{T. Giannini et al.}
   \maketitle
%

\section{Introduction}{\label{sec:sec1}
 
Photometric and spectroscopic variability is a common and defining feature of 
Young Stellar Objects (YSOs). When it occurrs on a short time-scale (hours or days), it is often related to flaring activity or to the formation
of hot or cool spots onto the stellar surface (e.g. Herbst et al. 1994), while
variability on longer time-scale (years or tens of years) is likely due to accretion bursts or to 
sudden changes in the local extinction.

Long-term, pre-main sequence variables are classified on the basis of the phenomenology of the  fluctuations, such as amplitude, duration,
recurrence time between subsequent events, presence of absorption or emission lines in the source spectrum (for a review see Audard et al. 2014).
Three phenomenological classes have been identified: FUor (Hartmann \& Kenyon 1985), EXor (Herbig 1989), 
and UXor (Herbst et al. 1994). The interaction between the  
circumstellar disk and the central star plays a fundamental role in all three 
classes: FUor and EXor bursts are caused by rapid variations of the mass accretion rate
from the disk, UXor events are due the recurrent appearance of an obscuring obstacle
likely located in the disk itself. Given the strong relationship with a non-spherically
symmetric structure, the emerging flux is expected to be largely influenced by the
inclination of the line of sight to the system. In some cases, this kind of circumstance 
might significantly affect the proposed scenario or even the object classification.

In this respect, V1184 Tau ($\alpha_{J2000.0}$= 05$^{h}$ 47$^{m}$ 03${\farcs}$77, $\delta_{J2000.0}$=
$+$21$^{\circ}$ 00$^{\prime}$ 34${\farcs}$79) is interesting because of its unusual photometric behavior.  
The source, likely seen edge-on, is a G5 III-IV  pre-main-sequence star, with estimated mass $\approx$ 2 M$_{\odot}$ and bolometric luminosity between 7 and 39 L$_{\odot}$ (Alves et al.1997). The object is seen in projection in the Bok globule CB 34 (Clemens \& Barvainis 1988), whose estimated distance is about 1.5 kpc. 
The intense star formation activity that occurrs in this
large globule was quantified using infrared (IR) and radio observations by Khanzadyan 
et al. (2002), who complemented the original discovery of Herbig-Haro (HH) objects 
(Moreira \& Yun, 1995).

The star  V1184 Tau attracted the attention of Alves et al. (1997)
and Yun et al. (1997), who both noted a brightening 
of about 4 mag between the Palomar Observatory Sky Survey (POSS) exposures of 1951 and 1994. Their original classification
of V1184 Tau as a FUor system has been reconsidered in the light of more recent observational studies. 
Its recent history has been monitored by Semkov (2004a, 2004b, 2006) and by Semkov et al. (2008, 2015) who
rejected the FUor original hypothesis and classified 
V1184 Tau as an UXor object that exhibits extinction-driven variations.  
Indeed, until 2004, the star presented  quite a constant flux level (I $\sim$ 13.0 mag), 
followed by a slow but deep fading (from 2004 to 2006, called the {\it '2004 photometric catastrophe'}) down to I $\sim$ 17 mag, overlapped by 
short-time variability  episodes (with amplitude of up to 0.6-0.8 mag in the J band), which  occur on timescales of days. 
Such rapid variations  have been interpreted by Tackett et al. (2003) as due to the rapid formation of thermal spots onto the stellar surface.

Near-infrared photometry of V1184 Tau was presented by Grinin et al. (2009), who 
proposed a very interesting and comprehensive model, emphasizing the importance of the disk inner rims in both 
extinguishing and reflecting the star emission. According to the proposed scenario, the source variability is explained 
by the intermittent increase of the geometric thickness of the internal disk, which is caused by the migration of  material from the external disk toward the star.
Since the system is seen almost edge-on, the observational effect is to shield the central star, which is therefore seen at a low
level of brightness. The initial conditions are supposed to be restored by means of a burst in the accretion rate, which provokes the so-called deflation of the disk's internal rim 
and the clearing of the central star and which, therefore, appears at its maximum level of brightness. 

In the Grinin scenario, V1184 Tau is therefore an accretion-driven variable whose light curve closely resembles that of a UXor star. However, the subsequent literature does not 
report any increase of luminosity, a circumstance that leads Herbig 
(2008) to rule out the EXor hypothesis. He presents high-resolution 
optical spectroscopy when the source level was down to R $\sim$ 17.4 mag.  

To better clarify the nature of V1184 Tau, and in particular to test the Grinin's model, we have collected optical/near-IR
data in the period 2008$-$2015, when the star has returned to the brightness level it had before 2004 (R $\sim$ 14 mag). 
These data are presented in Section\,2, then analyzed and discussed, together with previous literature data in Section\,3. Our concluding remarks are given in Section\,4. 


\begin{table*}
\caption{Near-infrared photometry.\label{tab:tab1}}
\centering
\footnotesize
\begin{tabular}{ccccc|ccccc}
\hline\\[-5pt]
Date    & MJD & J & H & K  &  Date  &  MJD  & J  & H  & K \\ 
(yy mm dd)    &           &  \multicolumn{3}{c|}{(mag)}&  (yy mm dd) &            &\multicolumn{3}{c}{(mag)} \\
\hline\\[-5pt] 
2008 Sep 09  & 54716 &  14.69  &  12.38  &  10.54 & 2013 Oct 12 & 56577 &  11.58  &  10.59  &   9.95 \\  
2008 Oct 18  & 54757 &  14.82  &  12.45  &  10.58 & 2013 Oct 14 & 56579 &  11.33  &  10.35  &   9.77 \\  
2008 Oct 20  & 54759 &  14.70  &  12.44  &  10.58 & 2013 Oct 25 & 56590 &  12.33  &  11.13  &  10.24 \\  
2008 Oct 23  & 54762 &  14.80  &  12.45  &  10.52 & 2013 Oct 26 & 56591 &  11.58  &  10.54  &   9.96 \\  
2008 Nov 02  & 54772 &  14.64  &  12.40  &  10.54 & 2013 Oct 29 & 56594 &  11.35  &  10.48  &   9.81 \\  
2008 Nov 11  & 54779 &  14.70  &  12.49  &  10.63 & 2013 Dec 18 & 56644 &  11.46  &  10.55  &  10.04 \\  
2008 Nov 12  & 54782 &  14.85  &  12.52  &  10.67 & 2013 Dec 24 & 56650 &  11.38  &  10.44  &   9.99 \\  
2008 Nov 20  & 54790 &  14.75  &  12.63  &  10.78 & 2014 Feb 25 & 56713 &  12.99  &  11.28  &  10.17 \\  
2009 Feb 02  & 54884 &  14.40  &  12.19  &  10.41 & 2014 Feb 26 & 56714 &  13.01  &  11.35  &  10.26 \\  
2009 Feb 26  & 54888 &  15.09  &  12.06  &  10.28 & 2014 Mar 08 & 56724 &  12.67  &  11.13  &  10.10 \\  
2009 Feb 27  & 54889 &  14.39  &  12.15  &  10.38 & 2014 Mar 12 & 56728 &  12.63  &  11.22  &  10.24 \\  
2009 Feb 28  & 54890 &  14.41  &  12.17  &  10.38 & 2014 Mar 13 & 56729 &  13.16  &  11.59  &  10.44 \\  
2009 Mar 01  & 54891 &  14.32  &  12.10  &  10.35 & 2014 Mar 14 & 56730 &  12.73  &  11.24  &  10.25 \\  
2009 Mar 09  & 54899 &  13.42  &  12.45  &  10.46 & 2014 Mar 15 & 56731 &  12.73  &  11.21  &  10.18 \\  
2009 Mar 14  & 54904 &  14.34  &  12.26  &  10.56 & 2014 Mar 17 & 56733 &  12.59  &  11.16  &  10.20 \\  
2009 Mar 17  & 54907 &  14.19  &  12.08  &  10.42 & 2014 Mar 18 & 56734 &  12.70  &  11.22  &  10.28 \\  
2010 Nov 03  & 55503 &  13.68  &  11.75  &  10.27 & 2014 Mar 19 & 56735 &  12.57  &  11.10  &  10.15 \\  
2010 Nov 04  & 55504 &  13.83  &  11.79  &  10.26 & 2014 Mar 20 & 56736 &  12.20  &  10.88  &  10.02 \\  
2010 Nov 06  & 55506 &  13.88  &  11.81  &  10.31 & 2014 Mar 21 & 56737 &  11.93  &  10.66  &   9.91 \\  
2010 Nov 13  & 55513 &  13.74  &  11.74  &  10.31 & 2014 Sep 30 & 56930 &  11.41  &  10.51  &  10.02 \\  
2010 Nov 14  & 55514 &  13.86  &  11.82  &  10.36 & 2014 Oct 09 & 56939 &  11.55  &  10.61  &  10.06 \\  
2010 Nov 15  & 55515 &  14.16  &  12.03  &  10.40 & 2014 Oct 20 & 56950 &  11.48  &  10.50  &   9.95 \\  
2011 Aug 23  & 55796 &  14.28  &  12.76  &  11.12 & 2014 Oct 29 & 56959 &  11.62  &  10.58  &  10.16 \\  
2011 Aug 25  & 55798 &  14.18  &  12.68  &  10.99 & 2014 Nov 02 & 56963 &  11.50  &  10.48  &   9.92 \\  
2011 Aug 29  & 55802 &  14.29  &  12.69  &  11.01 & 2014 Dec 24 & 57015 &  11.99  &  10.90  &  10.20 \\  
2013 Sep 27  & 56562 &  11.67  &  10.66  &  10.05 & 2015 Jan 16 & 57038 &  11.43  &  10.53  &  10.00 \\   
             &            &         &         &        & 2015 Aug 23\tablefootmark{a} & 57257  &  11.48  &  10.64  &  10.16 \\
\hline
\end{tabular}
\tablefoot{Typical errors of the near-IR magnitudes do not exceed 0.03 mag.
\tablefoottext{a}{Photometry taken with NICS at the {\it Telescopio Nazioale Galileo} (TNG).}}
\end{table*}

\section{Observations}{\label{sec:sec2}

New photometric and spectroscopic data of V1184 Tau were
obtained in the framework of our monitoring program of EXor dubbed EXORCISM (EXOR OptiCal 
and InfraRed Systematic Monitoring $-$ Antoniucci et al. 2013; Antoniucci et al. 2014; Lorenzetti et al. 2009).

\subsection{Photometry}{\label{sec:sec2.1}
\subsubsection{Near-IR photometry}{\label{sec:sec2.1.1}
Photometric images in the {\it JHK} broad-band filters were obtained during the period 2008$-$2015 
with the SWIRCAM camera (D'Alessio et al. 2000)
at the 1.1m AZT-24 telescope at the Campo Imperatore observatory (L'Aquila - Italy). An additional 
photometric point was also obtained with the NICS camera mounted on the Telescopio Nazionale Galileo (TNG)
on August 23 2015. All the observations were obtained by dithering 
the telescope around the pointed position. The raw imaging data were reduced by using standard 
procedures for bad pixel removal, flat fielding, and sky subtraction. Photometric data are listed 
in Table~\ref{tab:tab1} while the derived light curves are depicted in the three bottom panels of 
Figure~\ref{fig:fig1} for the $J$, $H$, and $K$ band, respectively (red points).

\subsubsection{Literature photometry}{\label{sec:sec2.1.2}
Near-infrared $JHK$ photometry prior to 2008 was obtained in 1993 by Yun et al. (1997), in 1997 (published in the $2MASS$ catalog), and between 2004 and 2008 by Grinin et al. (2009). These data 
are relevant for the present discussion since they clearly indicate that V1184 Tau is now in the 
same high state as it was before the deep fading that occurred between 2004 and 2009. These 
data in the literature are depicted in Figure~\ref{fig:fig1} as black points.

Optical photometry in $BVRI  $ that was obtained between 2002 and 2007 have been reported by Yun et al. 1997, Semkov 2003, 2004a, 2004b, 2006, Semkov et al. 2008, and by Grinin et al. 2009. Photometry in the period 2008$-$2015 (mainly in $V$ and $I$ bands) is available from the American Association of Variable Star Observers (AAVSO) webpage (www.aavso.org/) and recently published by Semkov et al. (2015).  Further photometric points were obtained by us together with the optical spectra (see Sect.\,\ref{sec:sec2.2}).
The optical light curves are plotted in Figure~\ref{fig:fig1}, from top to fourth panel.

Photometry at longer wavelengths was retrieved from the literature and from public databases. These data are reported in Table\,\ref{tab:tab2}
and cover the mid-IR domain from 3.4\,$\mu$m up to 24\,$\mu$m ({\it MSX}, {\it Spitzer} and {\it WISE} data). 


\begin{table*}
\footnotesize
\caption{\label{tab:tab2} Mid-infrared photometry}
\begin{center}
\begin{tabular}{ccccc}
\hline\\[-5pt]
$\lambda$  ($\mu$m)       &  Date                         & MJD            &   Photometry\tablefootmark{a}      &  Filter         \\
\hline\\[-5pt]
 3.4                      &  2009 Dec 14 - 2010 Aug 06    & 55180 - 55415  &    8.67$\pm$0.09      &  WISE3.4        \\
 3.4                      &  2010 Aug 06 - 2010 Sep 29    & 55415 - 55469  &    8.63$\pm$0.11      &  WISE3.4        \\
 3.6\tablefootmark{b}     &  2004 Mar 08 - 2004 Mar 15    & 53073 - 53080  &   7.87$\pm$0.01       &  Spitzer/IRAC3.6\\
 4.5\tablefootmark{b}     &  2004 Mar 08 - 2004 Mar 15    & 53073 - 53080  &   7.16$\pm$0.01       &  Spitzer/IRAC4.5\\
 4.6                      &  2009 Dec 14 - 2010 Aug 06    & 55180 - 55415  &    7.33$\pm$0.07      &  WISE4.6        \\
 4.6                      &  2010 Aug 06 - 2010 Sep 29    & 55415 - 55469  &    7.39$\pm$0.07      &  WISE4.6        \\
 5.8\tablefootmark{b}     &  2004 Mar 08 - 2004 Mar 15    & 53073 - 53080  &   6.55$\pm$0.01       &  Spitzer/IRAC5.8\\
 8.0\tablefootmark{b}     &  2004 Mar 08 - 2004 Mar 15    & 53073 - 53080  &   5.70$\pm$0.01       &  Spitzer/IRAC8.0\\
 8.3                      &  1996 - 1997                  & 50084 - 50814  &  0.37$\pm$0.03 (Jy)   &  MSX/A          \\
 12.0                     &  2009 Dec 14 - 2010 Aug 06    & 55180 - 55415  &    4.60$\pm$0.03      &  WISE12         \\
 12.0                     &  2010 Aug 06 - 2010 Sep 29    & 55415 - 55469  &    4.7$\pm$0.1        &  WISE12         \\
 12.1                     &  1996 - 1997                  & 50084 - 50814  &    $<$0.98 (Jy)       &  MSX/C          \\
 14.6                     &  1996 - 1997                  & 50084 - 50814  &    $<$1.1 (Jy)        &  MSX/D          \\
 22.0                     &  2009 Dec 14 - 2010 Aug 06    & 55180 - 55415  &    2.28$\pm$0.05      &  WISE22         \\
 24.0                     &  2004 Mar 08 - 2004 Mar 15    & 53073 - 53080  &    2.60$\pm$0.01      &  Spitzer/MIPS24 \\
\hline
\end{tabular}
\end{center}
\tablefoot{
\tablefoottext{a}{Photometry is given in magnitudes unless otherwise specified.}
\tablefoottext{b}{Taken from Gutermuth et al. 2009.}
 }
\end{table*}

\begin{figure*}
\centering
\includegraphics[width=14cm]{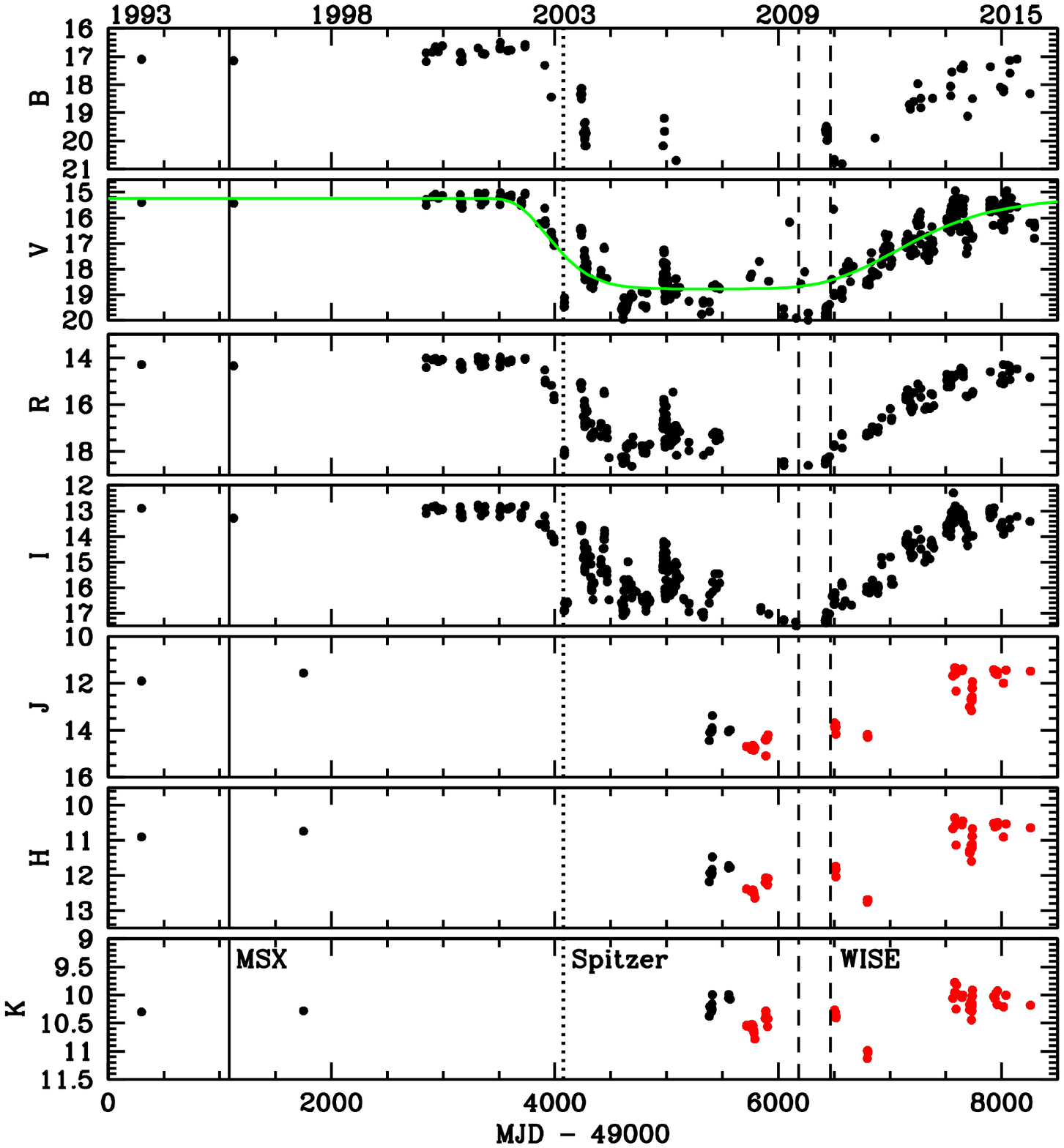}
\caption{V1184 Tau optical and near-IR light curves vs. Modified Julian Date (MJD). Red points are new infrared observations, while black points are literature data. Vertical black lines indicate the periods when
 mid-infrared data have been recorded. The corresponding telescope is also labeled. In the second panel we show, as an example, the fit through the $V$ light curve.
 The maximum and minimum values of the fit are assumed to be  $V$ magnitude of the low- and high-state of the source, and used to construct the corresponding SED (see Section\,\ref{sec:sec3.2}). \label{fig:fig1}}
\end{figure*}

\subsection{Spectroscopy}{\label{sec:sec2.2}

Optical spectra were taken on  December 12 2014 (MJD 57003) and August 19 2015 (MJD 57253). In addition, a NIR spectrum was 
obtained on August 23 2015 (MJD 57257).
The first optical spectrum (3\,200 \AA\, $-$ 10\,000 \AA) was obtained  with the 8.4m Large Binocular Telescope (LBT) 
using the Multi-Object Double Spectrograph (MODS - Pogge et al.
2010). The dual grating mode (Blue + Red channels) was used for a total integration time of 20 min 
to cover the spectral range with a 0.6\arcsec\, slit ($\Re \sim$ 2000). 
The second optical spectrum was obtained with the 3.6 m Telescopio Nazionale Galileo (TNG) with the
Device Optimized for the LOw RESolution (DOLORES) spectrograph. We adopted the low resolution ($\Re \sim$ 700)
red (LR-R) grism integrating 30 minutes to cover the 0.50-0.95 $\mu$m spectral range.
The images were bias and flat-field corrected using standard procedures. 
After removing sky background, the two-dimensional spectra were extracted and collapsed into one 
dimension. Wavelength calibration was achieved through lamp exposures, while flux
calibration was obtained from observations of spectro-photometric standard stars. 
The MODS spectrum 
of V1184 Tau is depicted (in different colors for each channel) in Figure~\ref{fig:fig2}.
The spectrum shows a number of photospheric absorption lines along with some fine structure atomic lines in emission.
In addition, we detect the H$\alpha$ line and the \caii$\lambda\lambda$ 8498, 8542, 8862  in absorption. 
Fluxes and equivalent widths (EW) of the main features are listed in Table\,\ref{tab:tab3}.

The integrated flux was therefore
obtained through a Gaussian fit to the (unresolved) line profile, while the associated uncertainty was derived 
by multiplying the {\it rms}  noise of the spectrum close to the line multiplied by the instrumental FWHM. 
A rough calibration in velocity was done by shifting the spectrum
 in wavelength so that the observed wavelength of  bright  photospheric 
lines (\Lii\, 6707\AA\, and \Ki\, 7699\AA\,) coincides with the theoretical air-wavelength value.
We estimate the velocity uncertainty to be $\sim$ 50 km s$^{-1}$. 

The DOLORES spectrum presents only three emission lines (H$\alpha$, [\oi]6300\AA\,, and [\sii]6730\AA\,), which  do not appear  to be significantly different
from those obtained with MODS (see Table\,\ref{tab:tab3}).

The near-infrared spectrum ($\Re \sim$ 500, slit width = 1$\arcsec$) was obtained with the Near Infrared Camera Spectrometer (NICS)  at TNG with two
grisms IJ (0.90 $-$ 1.4\,$\mu$m) and HK  (1.40 $-$ 2.50\,$\mu$m), each integrated for 25 minutes. For technical problems, we could not observe a telluric standard
star, which is fundamental to removing the atmospheric response. Therefore we could only examine the spectrum at the wavelengths where no atmospheric 
lines are present, and we do not detect any line that is intrinsic to V1184 Tau.


\begin{table}[h!]
\caption{\label{tab:tab3} Relevant lines detected on V1184 Tau with LBT/MODS.} 
\begin{center}
\begin{tabular}{cccc}
\hline\\[-5pt]
$\lambda_{air}$              & Line ID        &  Flux$\pm\Delta$(Flux)              &  EW    \\
  (\AA)                      &                &  (10$^{-15}$ erg s$^{-1}$ cm$^{-2}$)&  (\AA) \\
\hline
 5577.34                     &  [\oi]         &   1.1$\pm$0.15                      &  -0.8  \\ 
 6300.30                     &  [\oi]         &   3.9$\pm$0.3                       &  -1.5  \\ 
 6300.30\tablefootmark{a}    &  [\oi]         &   3.0$\pm$0.7                       &  -1.4  \\ 
 6363.78                     &  [\oi]         &   1.5$\pm$0.3                       &  -0.5  \\ 
 6562.80                     &  H$\alpha$     &  -0.8$\pm$0.2 / 8.9$\pm$0.3         &   0.2 / -2.8 \\
 6562.80\tablefootmark{a}    &  H$\alpha$     &   5.0$\pm$0.7                       &   -1.8 \\
 6583.46                     &  [\nii]        &   1.1$\pm$0.3                       &  -0.3  \\
 6730.81                     &  [\sii]        &   2.3$\pm$0.4                       &  -0.7  \\
 6730.81\tablefootmark{a}    &  [\sii]        &   1.5$\pm$0.4                       &  -0.5  \\ 
 8498.03                     &  \caii         &  -3.6$\pm$0.4                       &   0.6  \\
 8542.09                     &  \caii         & -10.9$\pm$0.5                       &   1.8  \\
 8662.14                     &  \caii         &  -9.1$\pm$0.5                       &   1.4  \\
 \hline
 \end{tabular}
\end{center}
\tablefoot{
\tablefoottext{a}{Line detected with TNG/DOLORES.}
}
\end{table} 
   
\begin{figure*}
\centering
\includegraphics[width=\hsize]{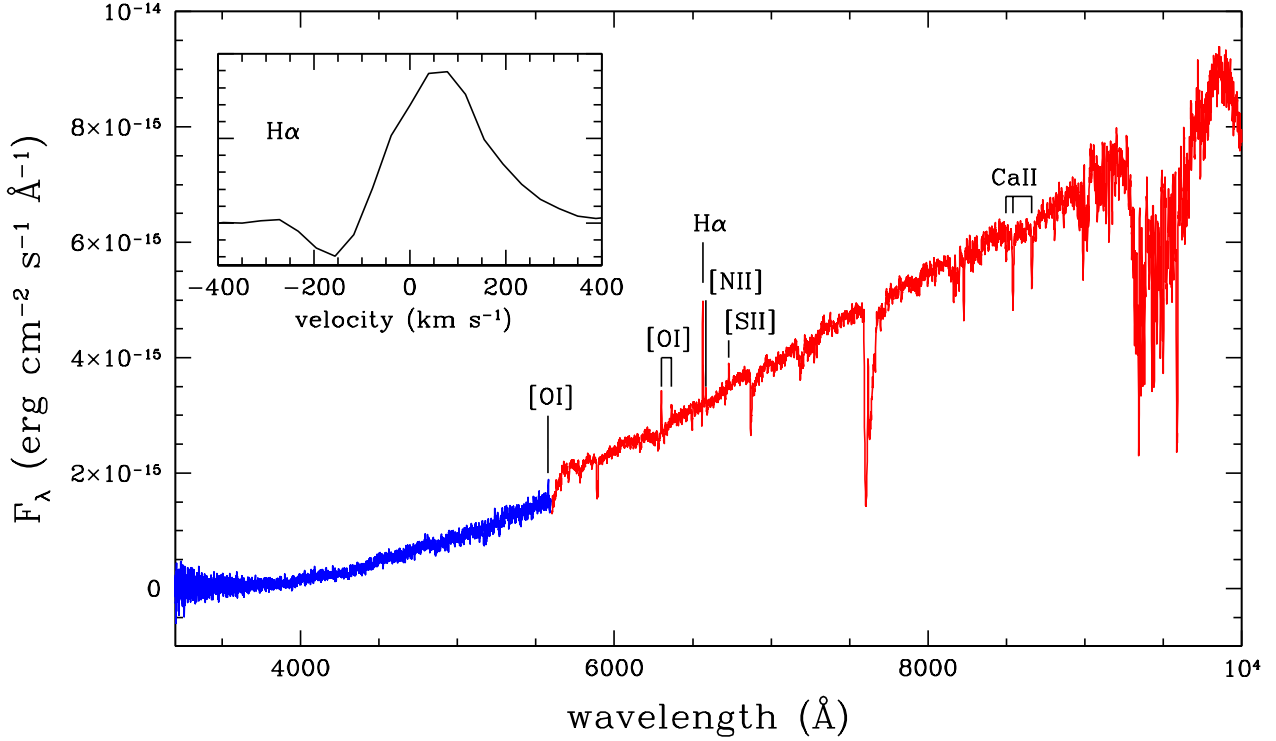}
\caption{LBT/MODS spectrum of V1184 Tau. Blue and red channels spectra are depicted with the corresponding color. Main features are labelled. In the insert we show the H$\alpha$ spectral profile.\label{fig:fig2}}
\end{figure*}
\section{Analysis and Discussion}{\label{sec:sec3}}

\subsection{Two-color plots}{\label{sec:sec3.1}}

\begin{figure*}
\includegraphics[width=\hsize]{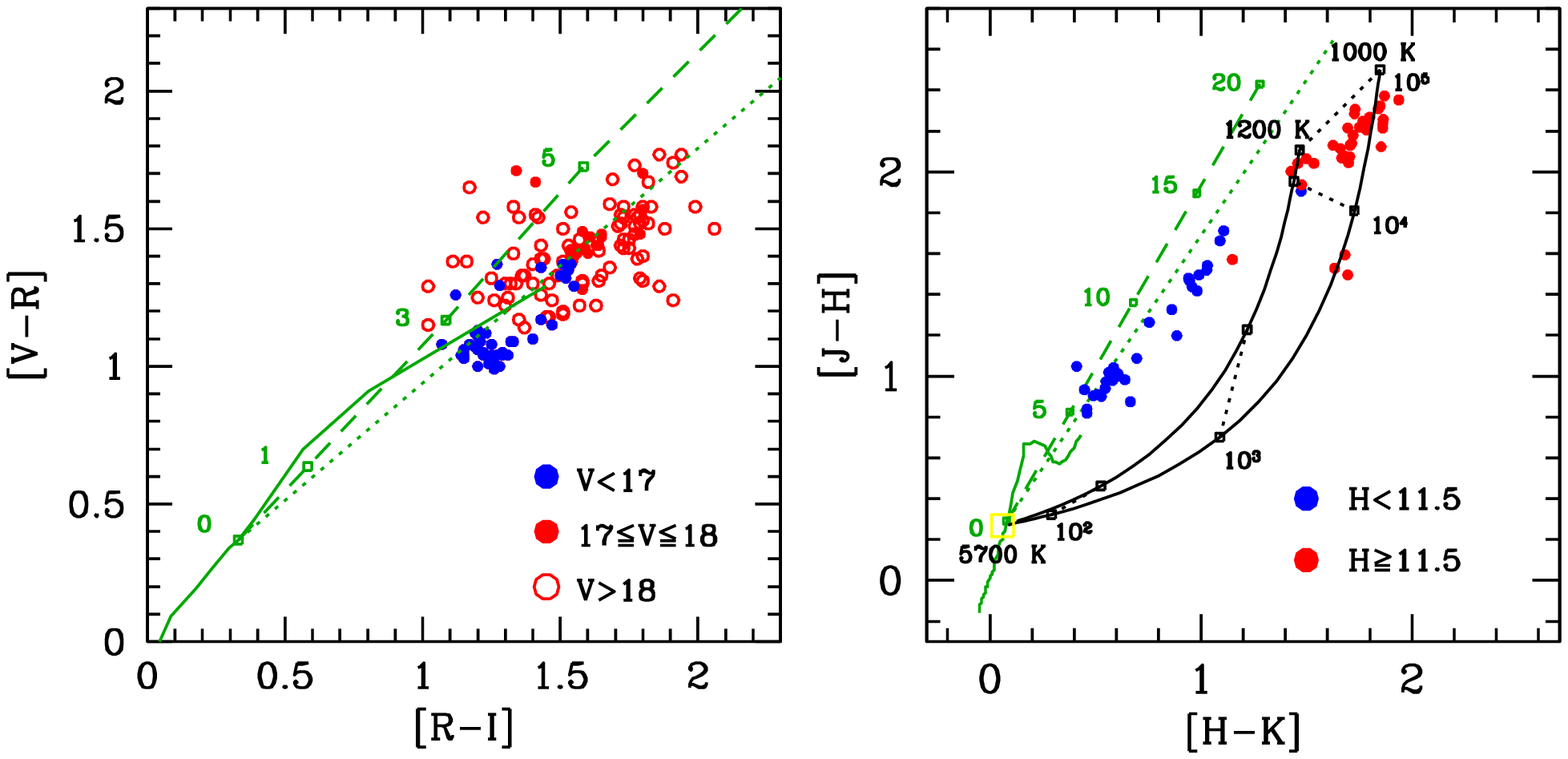}
\caption{{\it Left panel.} Optical $V$-$R$ vs. $R$-$I$ two-color diagram of V1184 Tau in different epochs. The solid green line represents the main sequence, while green dashed and dotted lines are the reddening vectors by Rieke \& Lebofsky (1985) and Cardelli et al. (1989), respectively. Different A$_V$ are indicated with open squares and labeled. The extinction vector is applied to the colors of a G5-type star. Blue dots show, individually, the states in which $V\,<$ 17 mag, red filled dots those with 17 $\le\,V\,\le$ 18 mag, and  red open
dots those with $V>$ 18 mag, respectively. {\it Right panel.} As in the left panel for the near-infrared $J$-$H$ vs. $H$-$K$ two-color diagram. Blue dots are the points with $H<$ 11.5 mag and red dots those with $H\ge$ 11.5 mag. The yellow square indicates the colors of the  photospheric temperature (5700 K), while black solid curves are the combinations of two blackbodies, one at 5700 K  and the second with temperatures between 1000 K and 2000 K. The ratio of the two emitting areas is labeled, being the iso-surface curves, depicted with black-dashed lines.
\label{fig:fig3}}
\end{figure*}

Optical ($VRI$ bands) and NIR ($JHK$ bands) diagrams of V1184 Tau are depicted in the two panels of 
Figure~\ref{fig:fig3}.  Together with the photometric data, we plot the {\it locus} of the main-sequence and two extinction vectors (Rieke \& Lebofsky, 1985 and  Cardelli et al., 1989), which were applied at the location
of the colors of a G5-type star.

Both plots indicate that the source is bluer when brighter, a typical behavior of all the accretion- and extinction-driven fluctuations (Lorenzetti et al. 2012). 

To go into  more detail: the optical two-color plot shows that data with $V \le$ 17 mag (blue dots) and with 17 mag $\le V \le$ 18 mag (red filled dots) tend to align along the 
extinction vector of Cardelli et al. (1989), and cluster at  A$_V$\,$\sim$\, 3 mag and A$_V$\,$\sim$\, 5 mag, respectively.
Conversely, data points corresponding to  $ V> $ 18 mag (open red dots) do not follow any of the extinction vectors. This 
effect, already noted in the  $V$ vs. $V-I$ and $V$ vs. $V-R$ color-magnitude diagrams by Barsunova et al. (2006) and  Semkov et al. (2008), is likely due to the scattering of the stellar light by small dust particles, which tends to prevail against the direct light from the star when this latter becomes fainter. This effect  makes the optical color-color plot effectively unable to probe 
a further extinction variation, as testified by the fact that the plot indicates  $\Delta$A$_V \sim$ 2 mag, while the  $V$ light curve gives a fading of more than 5 mag (Figure~\ref{fig:fig1}).

In the right panel we show the near-infrared $J$-$H$ vs. $H$-$K$ two-color diagram. Two main conclusions can be drawn: 1) the  data points do not align with either of two depicted extinction curves  but rather exhibit a clear shift to the right that increases with the decrease in the source brightness; 2) if only due to extinction, their spread should imply a $\Delta$A$_V$ larger than 15 mag, incompatible with the value derived in the optical light curve. Instead, the points indicate the presence of a strong IR excess that we quantify by plotting the locus of the combination of two thermal components, one at the stellar photospheric temperature of 5700 K  (that of a G5 star) and the second ranging between 1000 and 2000 K, and for different ratios between the emitting areas. In the high state, the infrared emission is largely dominated by the stellar emission, while in the faint state is it accounted for by a thermal source between 1000 K and 1200 K with an effective emitting area $\ga$ 10$^4$ larger than the stellar surface. 

\subsection{Spectral energy distribution}{\label{sec:sec3.2}}

\begin{figure*}
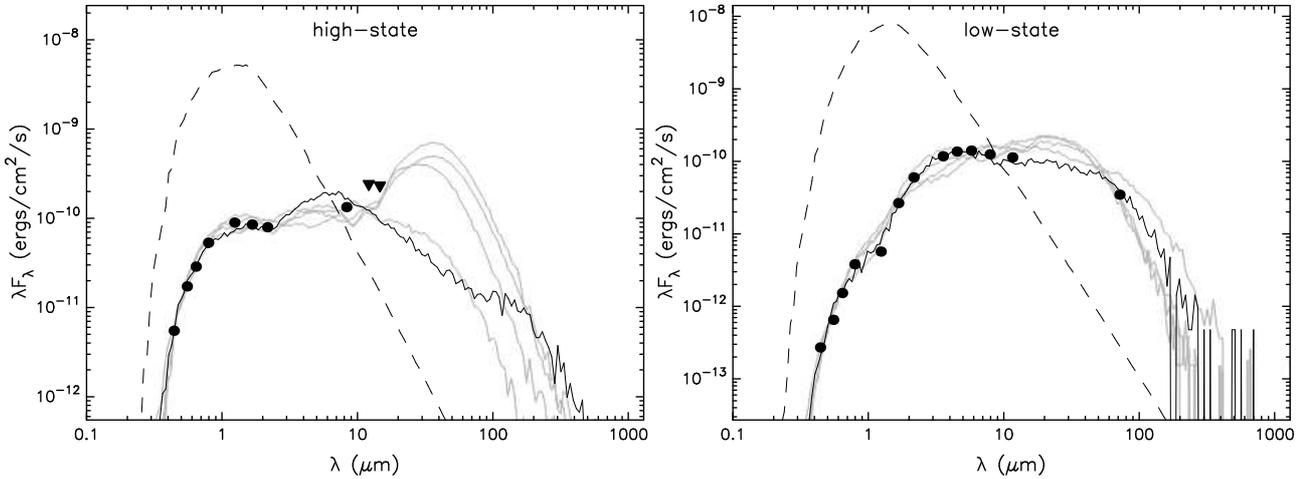

\includegraphics[width=8.5cm]{sed_high_150pc.eps}\includegraphics[width=8.5cm]{sed_low_150pc.eps}
\caption{{\it Left panel}: Fit "a la Robitaille", through the spectral energy distribution (SED) of V1184 Tau in the high-state.
The filled circles show the input fluxes, while triangles are 3-$\sigma$ upper limits. The black line shows the best fit, and the gray lines show four subsequent  good fits. The dashed line shows the stellar photosphere corresponding to the central source of the best-fitting model  (considering the interstellar extinction). {\it Right panel}: as in the left panel for the low state.
\label{fig:fig4}}
\end{figure*}
 

\begin{table}
\footnotesize
\caption{\label{tab:tab4} Output of the SED fitting}
\begin{center}
\begin{tabular}{ccc}
\hline\\[-5pt]
                                       & High state                & Low state        \\
\hline\\[-5pt]
D (pc)                                 &  {\bf 165}  (165-200)     & {\bf 151} (100-200)    \\
Inclination   ($\deg$)                 &  {\bf 87}   (87-87)       &  {\bf 87} (87-87)      \\
A$_V$   (mag)                          &  {\bf 2.3} (2.3-3.0)      & {\bf 4.0} (3.8-4.9)    \\
Age  (10$^6$ yr)                       &  {\bf 1.22} (0.5-7.8)     & {\bf 1.26} (0.9-7.0)   \\
T$_{\star}$ (K)                        &  {\bf 5171} (4800-7956)   & {\bf 5547} (5352-5933) \\   
M$_{\star}$ (M$_{\sun}$)               &  {\bf 3.10} (2.0-3.2)     & {\bf 3.37} (1.8-3.6)   \\
R$_{\star}$ (R$_{\sun}$)               &  {\bf 5.08} (1.89-5.73)   & {\bf 6.49} (2.45-6.87) \\
L$_{bol}$ (L$_{\sun}$)                 &  {\bf 16.7} (5.7-16.7)    & {\bf 35.7} (5.6-35.7)  \\
\.{M$_{acc}$} (10$^{-9}$M$_{\sun}$/yr) &  {\bf 4.2} (0.03-7.5)     & {\bf 0.2} (0.1-2.3)  \\
\hline
\end{tabular}
\end{center}
\tablefoot{The best-fit value is in boldface while, in parenthesis, is the range predicted by the first five best models.}
\end{table}

To construct the low- and high-state SEDs, we have first examined the optical and near-infrared light curves of Figure\,\ref{fig:fig1}. Here it is clear that a short-time variability, whose amplitude can exceed 0.6 $-$ 0.8 mag, superposes to the long-term variability in which we are interested. To integrate this short-time variability, we fitted all the data in a light-curve with a distorted Gaussian, whose maximum and minimum values are taken as the magnitude values typical of the low- and high-state. As example, we show in Figure\,\ref{fig:fig1} the fit through the $V$ light curve}. We applied this procedure to all the optical and near-infrared data.  In the mid-infrared, given the poor number of observations, we  simply attributed the measured fluxes to the high- or low-state, depending on the observation date. In particular, as evidenced in Figure\,\ref{fig:fig1}, 
both Spitzer and WISE observations have been performed during a relatively low-state period (we averaged the data taken at similar wavelengths i.e. 3.4/3.6 $\mu$m, 4.4/4.5 $\mu$m, and 22/24 $\mu$m), while the contrary happens for MSX observations for which, however, only that at 8.3 $\mu$m is a detection above the 3$-\sigma$ limit.  As a consequence, the high-state SED remains poorly defined.

The high- and low-state SEDs are shown in Figure\,\ref{fig:fig4}. We used the on-line SED fitter by Robitaille et al. (2007), which uses a Monte-Carlo radiation transfer code to compute model SEDs for different sets of physical parameters and viewing angles.
The free parameters are the distance and the visual extinction to the source. As a first attempt, we fitted
the data assuming a distance of between 0.5$-$2 kpc and A$_V$ between 0$-$10 mag. For consistency, we request that the fits of the two states do give similar outputs
for a number of parameters, such as viewing angle, stellar temperature, mass, and radius. However, we do not find this consistency within the first five fits (ordered by increasing $\chi^2$). Moreover, the 
predicted parameters are in clear contrast with those known for V1184 Tau. As an example, the fitted stellar temperature is between 8000 and 22000 K, the mass higher than 5 M$_{\sun}$, and the bolometric luminosity larger than 200 L$_{\sun}$. \\
More consistent results are found if the distance to the source is reduced. If this latter is fixed in the range 100$-$300 pc, we find output parameters in a much better agreement with  expectations. The 
first five model SEDs   (those with $\chi^2$ $\la$ 2$\chi^2_{bestfit}$)  are shown in 
Figure\,\ref{fig:fig4}. The main output parameters are listed in 
Table\,\ref{tab:tab4}, where we report the best-fit value (in boldface) together with the range predicted by  subsequent models. The two low- and high-state fits both predict the stellar 
parameters (temperature, age, bolometric luminosity, mass, and radius), which are in optimal agreement with literature estimates and the viewing angle is consistent with an object seen approximately edge-on. The mass accretion rate
is not well determined, but is roughly consistent with  predictions of normal T Tauri stars of  similar mass (Alcal\'a et al. 2014).\\  
The above scenario implies the distance to V1184 Tau to  be much lower than 1.5 kpc, as commonly accepted and the base of the distance of the Gemini OB1 cloud (Haug 1970), from which the CB34 globule is supposed to be generated. This estimate was put into doubt by  Khanzadyan et al. (2002), who note that the time needed to cover the distance (2$^\circ$ at 1.5 kpc) between CB34 and the Gemini OB1 cloud is comparable to the average age  for globule dispersal ($\sim$ 5 $\times$ 10$^6$ yr, Launhardt \& Henning 1997). Therefore, a direct estimate of the distance to CB34 is necessary to resolve the question. 

\subsection{Comments on the optical spectrum}{\label{sec:sec3.3}}
While the photometric behavior of V1184 Tau has been monitored for long time, little is known about the (possible) spectroscopic variability.
The first low-resolution spectra report the detection of  H$\alpha$ and [\oi]6300\AA\, (Alves et al. 1997;  Semkov 2003, 2004a, 2006). Later, a high-resolution spectrum was obtained by Herbig in 2005, i.e., when the star was at the lowest brightness level, showing a number of forbidden lines of \oi\,, \n\,, \sii\,, and \feii\,, together with H$\alpha$, H$\beta$, and \caii\, lines (Herbig 2008).

The comparison of our MODS-LBT spectrum with the previous ones leads to the following conclusions :
\begin{itemize}
\item[1] The EW of H$\alpha$ ($\sim$ 3\,\AA\,) is similar to that reported by Semkov (2006), which refers to observations taken on 2001 March, when the star was at a similar level of brightness (R\,=\,14.08) as in 2014 (R\,$\sim$\,14.3).
Conversely, in 2005 EW(H$\alpha$) was $\sim$90\AA\, and  R\,=\,17.38 (Herbig 2008). The simultaneous variation in EW and continuum level implies that the H$\alpha$  flux remains substantially unchanged.  Admittedly, expected UXor  behavior envisages a simultaneous decrease of
the continuum and the H$\alpha$ flux, since the circumstellar dust cloud screens both of them (e.g., Kolotilov 1977; Holtzman et al. 1986; Grinin et al. 1994). Conversely, an
increase of the EW(H$\alpha$)  with a fading of the stellar continuum can be explained  by: 1) an increase of the line luminosity; 2) a shielding of the star 
by a puffed-up inner disk rim that obscures just the stellar continuum, but leaves unaffected the hydrogen emission region (see the model by Dullemond et al. 2003), as observed in several UXor (Rodgers et al. 2002). This latter hypothesis seems more acceptable than 1), because in that case, to leave  the H$\alpha$ emission constant, the flux enhancement should  fortuitously compensate for  the effect of the screening of the hydrogen emitting zone.
\item[2] In the 2008 spectrum, the H$\alpha$ profile peaks at
about +60 km s$^{-1}$, with a shoulder (in emission) between -100 and -300 km s$^{-1}$ (Figure 7 of Herbig 2008). In our spectrum (insert in Figure\,\ref{fig:fig2}), this high-velocity component is seen as a faint 
absorption that peaks at velocity $\sim$ -150 km s$^{-1}$. In the assumption of a disk-wind origin of this kind of component, we can hypothesize that, if the star is shielded by a puffed-up disk rim, the bulk of the high-velocity component of the H$\alpha$ arises from the disk itself and, therefore, it is seen in emission while, when the line of sight is free from dust, the (warmer) wind that arises from the star is absorbed by the surrounding gas and, therefore, the line is seen in absorption. We note that, in agreement with the discussion above, the hypothesis of a passive dusty cloud in orbit around the star is not favored with respect to the 
the formation of a dusty, puffed-up inner disk rim, from which the wind should be generated. 
\item[3] A rough estimate of the mass accretion rate in the high state can be derived from the H$\alpha$ flux, although this line is not the best tracer for accretion (Antoniucci et al. 2011). Adopting the relationship given in Alcal{\'a} et al. (2014), and taking the range of parameters, obtained by fitting the SED , i.e. D = 100$-$200 pc, M$_{\star}$ $\approx$ 2$-$4 M$_{\odot}$, R$_{\star}$ $\approx$ 2$-$7 R$_{\odot}$ and  A$_V$ $\approx$ 2$-$3 mag (considering for this parameter the range fitted for the high-state SED) we get \.{M$_{acc}$} $\sim$  10$^{-11}$ $-$ 5 10$^{-10}$ M$_{\odot}$/yr, i.e. consistent with the lower end of the range that was  fitted with the Robitaille et al. model. More importantly, since the H$\alpha$ flux does not significantly change with the source continuum, it is likely that the mass accretion rate maintains a low value, even when the star is in its low-state phase.
\item[4] In our spectrum, we detect bright  \caii\, lines in absorption, while the same lines were seen in
 emission in the Herbig (2008) spectrum. 
At that time the velocity peak was $\approx$ -120 km s$^{-1}$ and EW $\sim$ 35 \AA, while we measure a peak velocity of a few km s$^{-1}$ and EW $<$ 2 \AA.\, As in the case of H$\alpha$,
a possible explanation is that the formation of a puffed-up rim is associated with that of a disk-wind that emits \caii\, lines. Conversely, when the rim deflates,  the disk wind also weakens and, therefore, the photospheric absorption component of the lines becomes visible. 
\item[5] The atomic forbidden line fluxes remain practically unchanged with respect to 2005. Their peak velocity is consistent with that measured by Herbig, and close to the cloud velocity. This suggests their origin in a nebular environment that should remain almost unperturbed by the phenomena that is responsible for both the stellar and permitted lines variation. \\

\end{itemize}
Summarizing the above, the scenario depicted from the optical lines substantially agrees with that proposed on the basis of the photometric behavior by Grinin (2009). In that paper, however, the mechanism that was causing the puffing of the inner disk rim is hypothesized as being the migration of the external disk material toward the central object, which should be followed by a sudden accretion burst and by a consequent deflation of the inner rim itself. Since the occurrence of an accretion burst is not supported by our spectroscopic observations, how the disk comes back to its original dimensions remains an open question.

\section{Conclusions}{\label{sec:sec4}}
We have presented new photometric and spectroscopic data that has been collected on V1184 Tau. Together with the literature data, these have been used to depict the phenomenology presented by the source in the last
 15 years.  
The main observational achievements of the presented study are: ({\it i}) multi-band photometry demonstrates that V1184 Tau has recently (end 2014) reached the same level of brightness that it had before the remarkable fading of about 5 mag started in 2004;
({\it ii}) the first optical spectrum, which was obtained in the high flux state, shows both similarities and differences in comparison with that taken at minimum brightness (Herbig, 2008). 

This type of scenario can be substantially explained based on the concept originally proposed by Grinin et al. (2009), but now supported by many more observational evidences. In particular, we can definitively discard the EXor nature of V1184 Tau on the base of: 1) the shape of the light curve(s), which does/do not present any sudden increase, but rather a smoother behavior: 2) the appearance of a strong IR excess that corresponds with the source fading, attributable to an emitting source with a dimension and temperature that is  compatible with the disk characteristics; 3) the roughly constant flux of H$\alpha$ and the estimated mass accretion rate, more typical of  slow accretors than outbursting variables. Consequently, all this evidence strongly favors V1184 Tau  being an UXor-type star. More specifically, the variation in both H$\alpha$ and \caii\, line profiles suggests a periodically broadening of the inner disk, which both enlarges enough to intercept the line of sight  (taking into consideration that the source is seen approximately edge-on) and originates  fast wind-emitting permitted lines. What still remains  unclear is whether or not the inner disk rim's so-called deflation is associated with accretion events (as envisaged by Grinin), given the negative clue that is offered by the roughly constant value
of the H$\alpha$ flux and, therefore, of the mass accretion rate.\\    
Although not directly connected with the above scenario is the result we got from the fit of the source SED, performed in the low- and high-state phases: the stellar parameters, independently determined in the literature, are correctly fitted only if the object is located at a distance that does not significantly exceed 200 pc, in contrast with the commonly accepted value of 1.5 kpc.  \\

\begin{acknowledgements}
This work is based on observations made with different instruments: [1] the Large Binocular Telescope (LBT). The LBT is an international collaboration among institutions in the United States, Italy, and Germany. LBT Corporation partners are: The University of Arizona on behalf of the Arizona university system; Istituto Nazionale di Astrofisica, Italy; LBT Beteiligungsgesellschaft, Germany, representing the Max-Planck Society, the Astrophysical Institute Potsdam, and Heidelberg University; The Ohio State University and The Research Corporation, on behalf of The University of Notre Dame, University of Minnesota and University of Virginia;
[2] the Italian Telescope Galileo (TNG), operated on the island of La Palma by the Fundaci\'{o}n Galileo Galilei of the INAF (Istituto Nazionale di Astrofisica) at the Spanish Observatorio del Roque
de los Muchachos of the Istituto de Astrofisica de Canarias;  [3] the AZT-24 IR Telescope at Campo Imperatore (L'Aquila - Italy), operated under the responsibility of the INAF-Osservatorio Astronomico di Roma (OAR)
according to the agreement between Pulkovo, Rome, and Teramo observatories. We thank the American Association for Variable Star Observers (AAVSO) for their optical monitoring of V1184 Tau. V.L. acknowledges support from  St. Petersburg University research grant 6.38.335.2015.
\end{acknowledgements}

{}

\end{document}